\documentclass[useAMS,usenatbib,usegraphicx]{mn2e}

\newcommand{\psrbl}{PSR~B1259-63/LS~2883}
\newcommand{\psrb}{PSR~B1259-63}

\title[Anisotropic inverse Compton scattering in \psrb]{Anisotropic inverse Compton scattering of photons from the circumstellar disc in \psrb\thanks{Based, in part, on observations made with ESO Telescopes at the Paranal Observatory under programme ID 086.D-0136(B), and on observations made with the Southern African Large Telescope (SALT) under program 2012-1-RSA-003. }}

\author[B. van Soelen, P.~J. Meintjes, A. Odendaal and L.J. Townsend]
  {B.~van~Soelen$^1$\thanks{E-mail:
vansoelenb@ufs.ac.za},
  P.~J.~Meintjes$^1$, A. Odendaal$^1$ and L.J. Townsend$^2$ \\$^1$Department of Physics, University of the Free State, South Africa, 9300\\
$^2$School of Physics and Astronomy, University of Southampton, Highfield, Southampton, SO17 1BJ}
\date{Released 2010 Xxxxx XX}

\begin{document}

\maketitle

\begin{abstract}
The gamma-ray binary system \psrbl\ consists of a 48~ms pulsar orbiting a Be star. The system is particularly interesting because it is the only gamma-ray binary system where the nature of the compact object is known.  The non-thermal radiation from the system is powered by the spin-down luminosity of the pulsar and the unpulsed radiation originates from the stand-off shock front which forms between the pulsar and stellar wind. The Be star/optical companion in the system produces an excess infrared flux from the associated circumstellar disc. This infrared excess provides an additional photon source for inverse Compton scattering.
We discuss the effects of the IR excess near periastron, for anisotropic inverse Compton scattering and associated gamma-ray production.  We determine the infrared excess from the circumstellar disc using a modified version of a curve of growth method, which takes into account the changing optical depth through the circumstellar disc during the orbit. The model is constrained using archive data and additional  mid-IR observations obtained with the VLT during January 2011. The inverse Compton scattering rate was calculated for three orientations of the circumstellar disc. The predicted gamma-ray light curves show that the disc contribution is a maximum around periastron and not around the disc crossing epoch. This is a result of the disc being brightest near the stellar surface.  Additional spectroscopic and near-infrared observations were obtained of the system and these are discussed in relation to the possibility of shock heating during the disc crossing epoch.  
\end{abstract}

\begin{keywords}
radiation mechanisms: non-thermal -- pulsars: individual:PSR~B1259-63 --X-rays: binaries
\end{keywords}

\section{Introduction}

The gamma-ray binary system \psrbl\ consists of a 48~ms pulsar orbiting around a Be star.  The orbit is eccentric  ($ e \approx 0.87 $) with an orbital period of  $P_\rmn{orb}\sim3.4$~yr \citep{johnston92,johnston94}.  The system is particularly interesting because it is the only gamma-ray binary system where the nature of the compact object is known.  Unpulsed TeV gamma-ray emission has previously been detected around periastron by the  High Energy Stereoscopic System (H.E.S.S.) \citep{aharonian05,aharonian09}.  Previous observations have also detected unpulsed radio emission near periastron \citep[e.g][]{johnston99} and  X-ray emission across the whole orbit \cite[see e.g.][]{cominsky94full,kaspi95,chernyakova06,chernyakova09}.  The discovery of extended radio and X-ray structures have been recently reported \citep*{moldon11, pavlov11}.  These observations have shown that the unpulsed emission is variable at all wavelengths with respect to the orbital period.  
The system was observed around the 2010 periastron passage ($\sim$ 14 December 2010) with  {\em Fermi}/LAT in the high energy (HE, $0.1-100$~GeV) regime and showed a flux modulation during the periastron passage, peaking $\sim 30$~days after periastron \citep{abdo11, tam11}. 

The non-thermal radiation from the system is powered by the spin-down luminosity of the pulsar and the unpulsed radiation originates from the stand-off shock front which forms between the pulsar and stellar wind.  Extensive modelling of this process has been presented in \citet{tavani97}, where the X-rays are produced via synchrotron radiation of ultra-relativistic electrons (Lorentz factor $\gamma \sim 10^6$) and gamma-rays through the inverse Compton (IC) scattering of target photons from the Be star.  The ratio between the X-ray and gamma-ray flux is dependent on the radiative and adiabatic cooling times of the post-shocked pulsar wind. IC cooling was predicted to increase near periastron.  IC cooling of the pre-shocked pulsar wind was also suggested for \psrb\ by \citet{ball00}.

The optical companion in the system, LS~2883,  is a Be star, and produces an infrared (IR) flux which is higher than predicted by Kurucz stellar atmosphere models \citep{kurucz79}.   This IR excess is produced by free-free and free-bound emission in the circumstellar disc which surrounds the Be star.  These Be star circumstellar discs are decretion discs which form around Be stars and are regions of enhanced stellar outflow.  Optical interferometry observations of these extended circumstellar envelopes have shown that they are symmetrical around the rotation axis \citep[see e.g.][]{quirrenbach94}.
The decretion discs are variable and grow and shrink over periods of hundreds to thousands of days.  See e.g.\ \citet{porter03} for a review of classical Be stars.  

The IR excess from the circumstellar disc provides an additional photon source to the blackbody flux from the optical star.  \citet{vansoelen11} showed that in the isotropic case this IR excess could increase the gamma-ray flux, particularly at GeV energies, by a factor $\sim 2$.   It is possible that near the disc crossing the IR excess may have a much greater effect.
In this study we discuss the effects of the IR excess near periastron, adopting an anisotropic inverse Compton model \citep*[e.g.][]{fargion97,dubus08}.  The soft photon excess provides target photons for IC scattering in the Thomson limit where the scattering cross section is a maximum. 
The additional effects of the different electron cooling processes in the plerion shock front \citep*[e.g.][]{kirk99}, as well as the effects of gamma-ray absorption \citep{dubus06}, will not be discussed in this paper and the reader is directed to the given reference for relevant discussions. In order to include the IR excess, we determine the target photon distribution using a modified version of the curve of growth method which was first proposed by \citet{waters86} (referred to hereafter as the COG method).  This modified COG model takes into account the changing optical depth as seen by the pulsar during its orbit.  We constrain the COG model using literature data and VLT observations which were taken during January 2011, approximately 22~days after the December 2010 periastron passage.

This paper is structured as follows: an outline of previous observations and modelling of \psrb\ is presented in section~\ref{psrb}; section~\ref{sec:observations} discusses observations we obtained with the VLT, SALT,  SAAO~1.9m~ and IRSF~telescopes after the 2010 periastron passage; section~\ref{be-stars} discusses the modified COG method that we have derived;  the anisotropic IC scattering is discussed in section~\ref{ic}; section~\ref{sec:modelling} outlines our modelling of the anisotropic inverse Compton scattering around periastron; sections~\ref{sec:discussion} and \ref{sec:conclusion} discuss the results and implications thereof.

\section{PSR B1259-63/LS 2883}
\label{psrb}

A few optical and IR photometric observations of LS~2883 are reported in literature and available in bright star catalogues.  A comparison of the catalogues' observations show little change in the optical magnitude.  
\citet{johnston94} confined the spectral range of LS~2883 to a O9 - B2 type star with a colour excess of $E(B-V)=1.05$ and assumed a spectral type B2e, with $M_* \simeq 10$~M$_{\odot}$ and $R_* \simeq 6$~R$_\odot$.   Analysis of spectroscopic observations also showed that  H$\beta$ emission occurred at a radius of $8.5 R_*$, assuming a Keplerian circumstellar disc, which implies that the  circumstellar disc has a radius $R_\rmn{disc}> 8.5 R_*$ \citep{johnston94}.  

More recently, \citet{negueruela11} undertook spectroscopic observations of LS~2883 and proposed new system parameters.  These authors estimate an O9.5~V star with a mass of 31~M$_\odot$ (for a non-rotating star), a radius of 9.2~R$_\odot$ and an effective temperature of 33\,500~K.  The authors also suggest a new binary inclination of $i=33$\degr\ and  a lower colour excess of $E(B-V) = 0.85$.

The size of the Be star's circumstellar disc can be constrained by the observed pulsar eclipse around periastron.   Observations of the system during consecutive periastron passages show that the pulsed radio signal is eclipsed  between roughly 20 days before periastron until 20 days after.  This is believed to be the result of the pulsar passing behind the Be star's circumstellar disc.  The binary separation at $\sim 20$ days is $\sim 45R_*$ (assuming $M_*=31~\rmn{M}_\odot$ and $R_*=9.2~\rmn{R}_\odot$) which implies that the disc radius is $R_\rmn{disc} \ga 45R_*$. For this reason we have adopted a disc radius of $R_\rmn{disc}=50R_*$ in this study.  
It should be noted, however, that Smooth Particle Hydrodynamic (SPH) simulations by \citet{okazaki11} and \citet{takata12} suggest that the circumstellar disc might be ``blown-away'' by the pulsar wind if the base density of the disc is low.  However, for larger densities, the disc remains more stable and these authors suggest that the density near the base of the disc is $\rho_0 \sim 10^{-9}$~g~cm$^{-3}$. This study only considers a constant disc size.

\psrb\ was detected by {\em Fermi}/LAT around the 2010 periastron with unexpected results.    
While the source remained faint before and just after periastron, approximately $30$~days after periastron, the source became very active and much brighter \citep{abdo11,tam11}. This was unexpected as it was assumed that the system would show fairly symmetric emission around periastron. Radio and X-ray observations around the same period showed no unusual or flaring events and were consistent with previous periastron passages \citep{abdo11}. Significantly, the peak luminosity detected by {\em Fermi}/LAT was $L_\gamma \approx 8\times 10^{35}$~erg~s$^{-1}$, which corresponds to approximately 100 per cent of the spin-down power of the pulsar.
No pulsed gamma-ray emission was detected with {\em Fermi}/LAT from \psrb\ during this period. This is unusual as \psrb\ fits most of the criteria normally associated with gamma-ray pulsars \citep{abdo11}.  The lack of a pulsed gamma-ray signal from a known radio pulsar poses interesting questions regarding the location of gamma-ray pulse production.  See e.g.\ \citet{abdo10_fermi_pulsar} for a discussion of the gamma-ray pulsars detected by {\em Fermi}/LAT.

Possible explanations for the {\em Fermi}/LAT flare have been suggested in terms of IC scattering of the cold pulsar wind \citep{khangulyan12}, or Doppler shifted synchrotron emission from the tail of the pulsar bow shock \citep{kong12}.

\section{Observations}
\label{sec:observations}

We obtained mid-IR observations of \psrbl\ with the VLT telescope during January 2011.  A number of observations were also undertaken at the SAAO\footnote{South African Astronomical Observatory} Sutherland observatory: we obtained near-IR observations with the IRSF telescope during December 2010, and spectroscopic observations with the SAAO~1.9m telescope and with SALT\footnote{Southern African Large Telescope} in April 2011 and May 2012 respectively.  The observations are summarised below.

\subsection{VLT observations}
\label{vlt}

We obtained (in Service Mode) mid-IR observations of \psrbl\ with the VLT Imager and Spectrometer of the mid Infrared \citep[VISIR;][]{lagage04}  with the VLT on 5 January 2011.  This is approximately three weeks after the 14 December 2010 periastron passage. The observations consisted of one pointing in multiple bands in the broad N-band region (8 -- 13~$\mu$m).   The observations were executed with a perpendicular chop/nod direction, a chopping amplitude of 10\arcsec\ and a random jitter of 3\arcsec.  The total integration time was of the order of 60~seconds per filter and achieved $S/N$ ratios of $\la 10$.  
The observations were calibrated with the standard star HD92682, which was observed as part of the standard ESO procedure, allowing for an average calibration accuracy within $\sim3$ per cent. \citep{smette07}.

The calibration star data were reduced with the \texttt{visir\_img\_phot} pipeline and the science target observations were reduced using the standard \texttt{visir\_img\_combine} pipeline to correct for the image nodding and chopping. Both reduction pipelines produce a single FITS \citep*{wells81} file with four measurements of the targets contained within each file. We calculated a flux conversion ratio for each filter by comparing the measured flux from the standard star to the defined standard flux. The flux conversion ratio was used to calibrate the measured flux from the science target.  

The final error in the VLT observation of LS~2883 was calculated by combining the estimated photometric error of each flux measurement and the standard deviation of the four measurements in each FITS file, for both the science target and standard star.  Not included in the error calculation is the $\sim 3\%$ error associated with the calibration process and any error in the standard star's catalogue flux value.    The calibrated  (but not de-reddened) flux at each filter is summarized in Table~\ref{tab:VLT}.

\begin{table}
\centering
\caption{VLT/VISIR data for LS~2883 observed on 5 January 2011.  Flux measurements were calibrated using the standard star HD92682,  but are not corrected of interstellar reddening.} 
\label{tab:VLT}
\begin{tabular}{lcc}\hline
Filter & central wavelength & Flux \\
	&	$\mu$m	&	mJy	\\	\hline		
PAH1	&	$8.59$	&	$261\pm40$	\\	
ArIII	&	$8.99$	&	$241\pm30$\\		
SIV	&	$10.49$	&	$218 \pm32$	\\	
SIV\_2	&	$10.77$	&	$188 \pm21$	\\	
PAH2	&	$11.25$	&	$208 \pm31$	\\	
SiC	&	$11.85$	&	$198 \pm18$	\\	
PAH2\_2	&	$11.88$	&	$196\pm 54$\\		
NeII	&	$12.81$	&	$143 \pm51$	\\	\hline
\end{tabular}
\end{table}

\subsection{SAAO observations}

\subsubsection{Spectroscopic observations}

\begin{figure}
 \centering
 \includegraphics[scale=1]{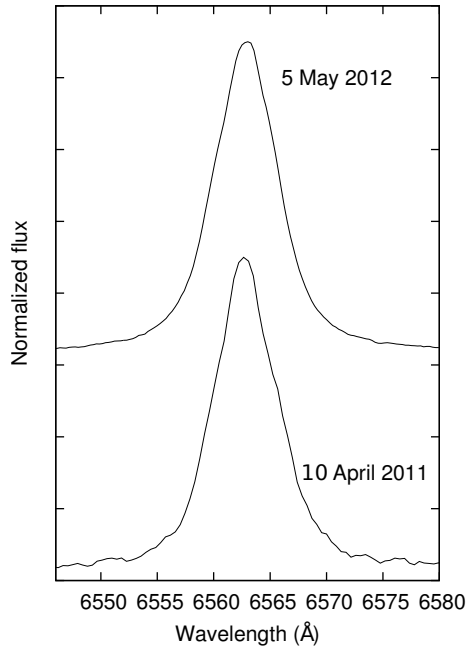}
 \caption{H$\alpha$ emission line of \psrbl\ taken 10~April 2011 with the SAAO~1.9m telescope and on 5~May 2012 with SALT. The wavelength range has been corrected to the heliocentre.}
 \label{fig:ls2883_halpha}
\end{figure}

We obtained spectroscopic observations of the H$\alpha$ emission line from \psrbl\ on 10 April 2011  with the SAAO~1.9m telescope and on  6 May 2012 with SALT\footnote{Program: 2012-1-RSA-003} (Fig.~\ref{fig:ls2883_halpha}).  The SAAO~1.9m observations were taken using a grating spectrograph (1200 lines $\mbox{mm}^{-1}$ grating) which gives a $\sim$ 1 \AA\ resolution.  The data was reduced and analysed  using the standard {\sc iraf/noao} packages.\footnote{{\sc iraf} is distributed by the National Optical Astronomy Observatories, which are operated by the Association of Universities for Research in Astronomy, Inc., under cooperative agreement with the National Science Foundation. See e.g.\ \citet{tody93}.} The SALT observations were performed with the 0.6\arcsec\ slit and the pg2300 grating, achieving a resolution of $R\sim10000$.  The SALT data was pre-reduced with the standard SALT pipeline \citep{crawford10} and the data analysis was done with the standard {\sc iraf/noao} packages.   The H${\alpha}$ emission line is prominent in both observations and there appears to be no indication of stellar H${\alpha}$ absorption.  This indicates that the circumstellar disc completely dominates the stellar absorption at the same frequency.  The observed equivalent width for the H$\alpha$ line was EW~$= -50\pm5$~\AA\ in April 2011 and $\rmn{EW} = -49\pm2$~\AA\ in May 2012.  This indicates that the Be star was still surrounded by a substantial circumstellar disc after the disc crossing epoch.   The observed equivalent width in April 2011 is in line with  the measurement reported by \citet{negueruela11} (EW~$=-54\pm2$~\AA), while the emission line is smaller in May 2012.  An equivalent width of $\rmn{EW} =40$~\AA\ was reported earlier by \citet{johnston94}, and the smaller equivalent width detected in May 2012 most likely represents the normal variability associated with Be stars.

\subsubsection{IRSF}

Observations were undertaken with the InfraRed Survey Facility (IRSF)  on 19 December 2010. The IRSF is a 1.4~m near-IR telescope, located at SAAO near Sutherland and was built as a collaboration between SAAO and the Nagoya University in Japan.   Using the SIRIUS\footnote{Simultaneous-3color InfraRed Imager for Unbiased Survey} camera, the IRSF takes simultaneous J, H and $\rmn{K_s}$ observations. The data was reduced using the standard IRSF data reduction pipeline and data from the \textit{2MASS} point-source catalogue \citep{2mass} was used to self-calibrated the frame. The observation data is summarized in Table~\ref{tab:irsf}.  The final error is calculated by combining the photometric error estimation, the measured error in the \textit{2MASS} data and the error in the self-calibration fit.  The data is in accordance with the measurements obtained in the \textit{2MASS} catalogue.

\begin{table}
 \caption{IRSF observations obtained during December 2010}
\label{tab:irsf}
\begin{center}
\begin{tabular}{lccc} \hline
MJD & J & H & $\rmn{K_s}$\\ \hline
55549 & $7.999\pm0.038$ & $7.584\pm0.045$ & $7.176\pm0.082$\\ \hline
\end{tabular}
\end{center}

\end{table}

\section{The Curve of Growth Method} 
\label{be-stars}

\citet{waters86} proposed a COG method to model the IR excess of Be stars, by determining the optical depth through the circumstellar disc. This model  is appropriate for stars where the discs are viewed face--on at distances which are much larger than the size of the disc.  Below we present a modified version of this COG method which we used to model the IR excess originating from the circumstellar disc as observed from the pulsar.

In the COG model it is assumed that the circumstellar disc has a half-opening angle $\theta_\rmn{disc}$, extends to a radius $R_{\rmn{disc}}$, and has a power-law density profile that decreases with distance as 
\begin{equation}
 \rho(r) = \rho_0 \left(\frac{r}{R_*}\right)^{-n} = \rho_0 \bar{r}^{-n}, 
 \label{eqn:rho_distance}
\end{equation}
where $\bar{r} \equiv r/R_*$ is the radial distance in units of the stellar radius ($R_*$). It can be shown that the optical depth through the disc is given by 
\[
 \tau = E_{\nu,\rmn{disc}} \int \bar{r}^{-2n} \, \rmn{d}l,
\]
where $E_{\nu, \rmn{disc}}$ is the optical depth parameter for the circumstellar disc given by \citep{waters86}
\[
 E_{\nu,\rmn{disc}} = X_\lambda X_{*d}.
\]
 Here $X_\lambda$ is a wavelength dependent term,
\begin{eqnarray}
 \nonumber\lefteqn{ X_\lambda= \lambda^2 \{ ( 1 - e^{-h\nu/kT_\rmn{disc}} )/(h\nu/kT_\rmn{disc})\}} \\ 
& &\times \{g(\nu,T_\rmn{disc}) + b(\nu,T_\rmn{disc})\} \label{eqn:x_lambda}
\end{eqnarray}
where $\lambda$ is the wavelength, $\nu$ is the frequency, $k$ is the Boltzmann constant, $T_\rmn{disc}$ is the disc temperature, and $g(\nu,T_\rmn{disc}) + b(\nu,T_\rmn{disc})$ is the sum of the free-free and free-bound gaunt factors.
The variable $X_{*,d}$, which is fit by the COG method, is given by
\[
 X_{*d} = 4.923 \times 10^{35} \overline{z^2} T_\rmn{disc}^{-3/2} \mu^{-2}  \varpi \rho_0^2 \left(\frac{R_*}{R_\odot}\right),
\]
where $\overline{z^2}$ is the mean of the squared atomic charge, $\mu$ the mean atomic weight (in units of proton mass) and  $\varpi$ is the ratio of the number of electrons to the number of ions.

 We calculated the gaunt factors in equation~(\ref{eqn:x_lambda}) following the method outlined in \citet{waters84}.  This allows us to determine gaunt factors for a larger wavelengths range than was included in the published tables. 

\begin{figure}
 \includegraphics[scale=.55]{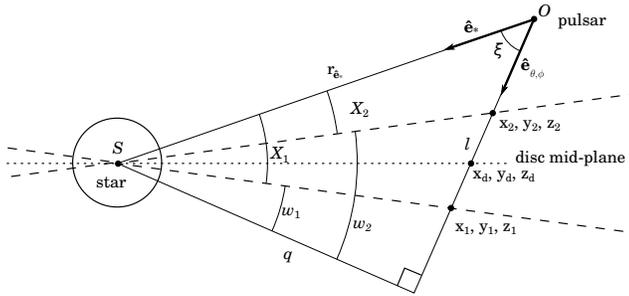}
 \caption{Line-of-sight through the circumstellar disc of a Be star. The centre of the pulsar/electron cloud is at $O$ and the centre of the star is at position $S$.  The optical depth along a line-of-sight through the disc is dependent on the length between the two disc intercepts $x_1, y_1, z_1$ and $x_2, y_2, z_2$. See text for details.}
 \label{fig:cog_fig_paper}
\end{figure}

The optical depth through the disc, as observed from any arbitrary point (along the direction $\bmath{\hat{e}_{\theta,\phi}}$ in Fig.~\ref{fig:cog_fig_paper}), can be defined in terms of an impact factor $q$, such that
\[
\bar{r}^2 = l^2 + q^2, 
\]
where all variables are in units of $R_*$.  Thus it follows that 
\[
 \tau = E_{\nu,\rmn{disc}} \int \bar{r}^{-2n + 1} \left(\bar{r}^2-q^2\right)^{-1/2} \, \rmn{d}\bar{r}.
\]
For any value of $q$, the substitution $w = \arccos\left(q/\bar{r}\right)$ can be made and it follows that
\begin{equation}
\tau(q) = E_{\nu,\rmn{disc}}\, q^{-2n+1} \int \cos^{2n-2} w \, \rmn{d}w, 
 \label{eqn:tau}
\end{equation}
where the integration is performed over suitable limits of $w$.  The limits on the angle $w$ are determined by the angle between the impact factor $q$ and the positions where the line-of-sight intercepts the disc. An example is shown in Fig.~\ref{fig:cog_fig_paper} where the limits are given by the angles $w_1$ and $w_2$.    For a disc observed face-on at infinity $q$ lies along the plane of the disc for all lines-of-sight and the limits on $w$ are $-\theta_\rmn{disc}$ to $\theta_\rmn{disc}$ \citep{waters86}. For an observer close to the disc the value of $\tau$ must be calculated differently for every line-of-sight as the optical depth will vary.  The calculation of the limits of $w$ and the value of $q$ are presented in Appendix~\ref{app:limits}.

It is assumed that the circumstellar disc is isothermal, and therefore the source function of the plasma in the disc can be described by Planck's function.  The specific intensity of the radiation field which originates from the disc is then determined by \citep[e.g.][]{wright75}
\begin{equation}
I_\nu = B_\nu(T_\rmn{disc}) \left( 1- \rmn{e}^{-\tau(q)} \right), 
\label{eqn:disc_intensity}
\end{equation}
where $B_\nu(T_\rmn{disc})$ is the Planck function, $T_\rmn{disc}$ is the temperature of the disc and $\tau(q)$ is determined by equation~(\ref{eqn:tau}).

The unmodified COG model, as presented by \cite{waters86} assumes that the circumstellar disc is viewed face--on, at a distance much greater than the size of the disc.  The impact factor, $q$, then lies along the mid-plane of the disc and has a maximum value of $q=R_\rmn{disc}/R_*$.  For $q >1$ the specific intensity is determined by equation~(\ref{eqn:disc_intensity}).  For $q \leq 1$ it is assumed the disc has a minimal influence on the intensity \citep{waters86} and then 
\[
 I_\nu = I_{\nu,*},
\]
where $I_{\nu,*}$ is determined from the appropriate Kurucz atmosphere or blackbody distribution. The ratio of the observed flux arising from the disc to the flux arising from the star is given by \citep{telting98}
\begin{equation}
 \frac{F_{\nu,\rmn{disc}}}{F_{\nu,*}} = \frac{B_\nu (T_\rmn{disc})}{I_{\nu,*}} \int_1^{R_\rmn{disc}/R_*} \left[1-e^{-\tau_\nu(q)}\right] \, 2q \, \rmn{d}q.
\label{equ:irratio}
\end{equation}
The equation above is appropriate to the flux which would be observed from Earth for a disc which is observed face-on.
The COG method can then be used to model a Be star's optical/IR flux by modifying a Kurucz model atmosphere at IR wavelengths with the flux ratio given by equation~(\ref{equ:irratio}), which is calculated by fitting the five parameters $n$, $X_*$, $R_\rmn{disc}$, $T_\rmn{disc}$ and $\theta_\rmn{disc}$ to IR data. 

\citet{waters86} showed that the observed flux from a moderately inclined disc, viewed from a very large distance, will not deviate greatly from that of a disc observed face-on.  For this reason we have calculated the parameters for the COG method using the unmodified model.  
However, when the pulsar is near the circumstellar disc, the disc target photon spectrum, as observed from the pulsar's position, cannot be calculated assuming the disc is viewed at infinity.  The optical depth, and hence the specific intensity and photon spectrum, must be calculated for all lines-of-sight through the disc from the position of the pulsar.  This will lead to a variability in the observed photon spectrum as the optical depth changes during the orbit. 

In the inverse Compton calculation we have calculated the photon density arising from the circumstellar disc by using the modified COG method outlined above.

\section{Inverse Compton Scattering}
\label{ic}

\subsection{Anisotropic scattering}

In order to model the effect of the circumstellar disc on IC gamma-ray emission in Be-XPBs, it is necessary to take into account the IR excess, the solid angle over the disc and the changing scattering angle. This  is achieved by calculating the anisotropic IC at all points in the orbit.  

Anisotropic scattering has been considered in detail by \citet{fargion97} and has been adapted to the case of LS~5039 by \citet{dubus08}.   The total scattering rate is given by
\begin{equation}
  \frac{\rmn{d}N_\rmn{tot}}{\rmn{d}t \rmn{d}\epsilon_1} = \int_{\Omega} \int_\gamma \int_{\epsilon_0} n_\rmn{ph}(\epsilon_0) n_\rmn{e}(\gamma) \frac{\rmn{d}N_{\gamma,\epsilon_0}}{\rmn{d}t\,\rmn{d}\epsilon_1} \cos \vartheta \,\rmn{d}\epsilon_0 \,\rmn{d}\gamma \,\rmn{d}\Omega, 
\label{eqn:AICtotal}
\end{equation}
where $n_\rmn{ph}$  and $n_\rmn{e}$ are the photon and electron number distribution respectively,
${\rmn{d}N_{\gamma,\epsilon_0}}/{\rmn{d}t\,\rmn{d}\epsilon_1}$ is the anisotropic scattering rate and the $\cos \vartheta$ term corrects for the projected surface area of the circumstellar disc.  The full equation for the Klein--Nishina scattering rate, ${\rmn{d}N_{\gamma,\epsilon_0}}/{\rmn{d}t\,\rmn{d}\epsilon_1}$,  is derived and given in \citet[][equation~A6 in their paper]{dubus08}.  The total scattering rate is calculated by integrating over the appropriate range of electron and photon energies and the solid angle subtending the target photon source.

\subsection{Integration over the solid angle}
\label{sec:angleofdisc}

The solid angle subtending the target photons (star and/or disc), from the pulsar's perspective, changes during the orbit.  This, combined with the change in the scattering angle, results in a modulation of the IC scattering rate.   For the case of the photon contribution from Be stars, the contribution from the disc must also be considered and the integration must occur over the whole disc.  This is accomplished by summing the scattering contribution from the star and the disc. 

The solid angle subtending the star  is given by
\[
 \rmn{d}\Omega = \sin \theta \, \rmn{d}\theta \rmn{d}\phi, 
\]
where $\phi \in [0,2\pi]$, $\theta \in [0,\alpha_*]$, and
\[
 \alpha_* = \arcsin\left( \frac{R_*}{r} \right).
\]
The size of the solid angle is then dependent on the binary separation, $r$,  between the star and the pulsar. 

\begin{figure}
 \includegraphics[scale=0.65]{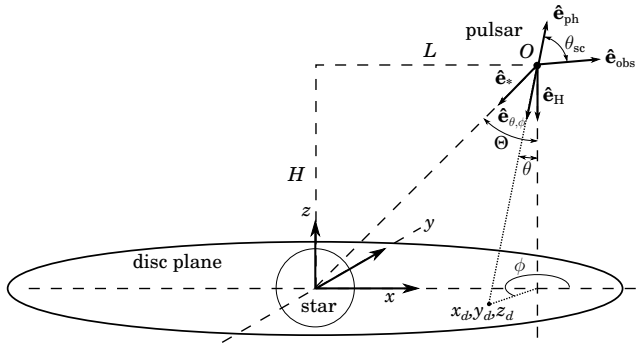}
 \caption{Coordinate system used to integrate over the solid angle subtended by a disc.}
 \label{fig:disc-geometry}
\end{figure}

In order to calculate the solid angle subtending the disc, the co-ordinate system shown in Fig.~\ref{fig:disc-geometry} is used. \citep[See e.g.][for a general discussion of the formulae below.]{pomme03,Moskalenko00,tryka97} 

The pulsar is at $O$ which is at a vertical height $H$ above the disc mid-plane and shifted a horizontal distance $L$ (along the $x$-axis) from the centre of the disc/star. 
The solid angle subtending the disc from the position $O$ can be calculated by dividing the integration into two regions, $\Omega_1$ and $\Omega_2$:  
\begin{eqnarray}
 \label{eqn:omega_one} \Omega_1 & = &\int_0^{2\pi} \int_0^{\theta_1} \sin \theta \, \rmn{d}\theta \, \rmn{d}\phi, \quad (L<R_\rmn{disc})\\
\nonumber & = & 0,   \quad ({L \geq R_\rmn{disc}),}
\end{eqnarray}
where
\[
 \theta_1 = \arctan \left( \left| \frac{R_\rmn{disc}-L}{H} \right| \right);
\]
and
\begin{equation}
 \Omega_2 =  \int_{\theta_1}^{\theta_2} \int_{\varphi(\theta)}^{2\pi - \varphi(\theta)} \sin \theta \,\rmn{d}\phi \, \rmn{d}\theta , 
\label{eqn:omega_two}
\end{equation}
where 
\[
 \theta_2 = \arctan \left( \frac{R_\rmn{disc}+L}{H} \right)
\]
and 
\[
 \varphi_{\phi} (\theta) = \arccos \left( \frac{R_\rmn{disc}^2 - L^2 - H^2 \tan^2\theta}{2LH\tan\theta}\right).
\]
The total solid angle of the disc is $\Omega = \Omega_1 + \Omega_2$.

The above equations apply to the solid angle over a flat disc ($\theta_\rmn{disc} = 0$\degr), but can easily be extended to a circumstellar disc with a finite half-opening angle $\theta_\rmn{disc}$.  The solid angle is given by equations~(\ref{eqn:omega_one}) \& (\ref{eqn:omega_two}) where the new limits on the polar and azimuthal angles are
\begin{eqnarray*}
 \theta_1 &=& \tan^{-1} \left(\left| \frac{L - R_\rmn{disc}}{H-h_\rmn{disc}}\right|\right), \\
 \theta_2 &=& \tan^{-1} \left( \frac{R_\rmn{disc} + L}{\left|H - h_\rmn{disc}\right|}\right), \\
 \varphi_\phi (\theta) &=& \cos^{-1} \left( \frac{R_\rmn{disc}^2 - L^2 - (H-h_\rmn{disc})^2 \tan^2\theta}{2L (|H-h_\rmn{disc}|)\tan\theta}\right).
\end{eqnarray*}
Here the additional term,  $h_\rmn{disc} = R_\rmn{disc} \tan \theta_\rmn{disc}$, accounts for the opening angle of the disc.

\subsection{Scattering angle}

The anisotropic IC (AIC) scattering rate is calculated by assuming that the electrons are highly relativistic and that the photons scatter in the direction of the electrons' original momentum.  This assumption was incorporated into the AIC scattering kernel \citep{dubus08}, and the scattering rate is dependent on the angle between the unperturbed path of the photon and the direction of the un-scattered electron, in the observer's reference frame.  This angle,  defined here as the scattering angle, $\theta_\rmn{sc}$,  is given by the angle between the direction of the unperturbed photons path ($\bmath{\hat{e}_{\theta,\phi}}$) and the direction of the observer ($\bmath{\hat{e}_\rmn{obs}}$), and is determined from (Fig.~\ref{fig:disc-geometry})
\[
\bmath{\hat{e}_\rmn{ph}} \cdot \bmath{\hat{e}_\rmn{obs}} =  \cos \theta_\rmn{sc}.  
\]

\section{Modelling the AIC scattering}
\label{sec:modelling}

To calculate the IC scattering rate the infrared excess from the circumstellar disc must first be determined.

\subsection{Modelling the infrared excess}
\label{sec:modelling_cog}

We fitted the un-modified COG model to available archive data and new VLT observations.  Archive optical data is available from a catalogue of bright stars near the Southern Coalsack region compiled by \citet{westerlund89}, while IR data is available from the \textit{2MASS} \citep{2mass}  and \textit{MSX}\footnote{Data from \textit{2MASS} and the \textit{Midcourse Space Experiment} (\textit{MSX}) are available on-line through the NASA/IPAC Infrared Science Archive.} \citep{price01} missions, as presented in \cite{vansoelen11}.
The combined VLT and archive data for LS~2883 were de-reddened with the {\sc dispo} software packages using the colour excess, E(B-V) = 0.85, proposed by \citet{negueruela11}.  The total combined and de-reddened data are shown in Table~\ref{tab:ls2883_all}.

\begin{table}   
 \centering   
\caption{Combined VLT and archive data for LS~2883 de-reddened using E(B-V)=0.85.}  
\label{tab:ls2883_all}
\begin{tabular}{lcc}\hline 
Band  &   Wavelength  &   $F_0$   \\
  &   $\mu$m  &   $10^{-24}$ erg/s/cm$^{2}$/Hz   \\ \hline
U &  0.365 &  $50.03\pm0.107$ \\
B  &  0.44 &  $52.433\pm0.088$\\
V &  0.55 &  $40.539\pm0.097$\\
J &  1.235 &  $19.614\pm0.066$\\
H &  1.662 &  $12.695\pm0.091$\\
K &  2.159 &  $10.677\pm0.029$\\
A &  8.23 &  $2.71\pm0.14$\\
C &  12.13 &  $10.91\pm5.88$\\
PAH1 &  8.59 &  $2.64\pm0.41$\\
ArIII &  8.99 &  $2.43\pm0.30$\\
SIV  &  10.49 &  $2.19\pm0.32$\\
SIV\_2 &  10.77 &  $1.89\pm0.22$\\
PAH2 &  11.25 &  $2.09\pm0.31$\\
SiC &  11.85 &  $1.99\pm0.18$\\
PAH2\_2 &  11.88 &  $1.97\pm0.54$\\
NeII &  12.81 &  $1.43\pm0.52$\\ \hline  
\end{tabular} 
\end{table}

Spectroscopic observations undertaken by  \citet{negueruela11} suggested a stellar temperature of $T_* = 33\,500$~K and an effective gravity $\log g = 4.0$ for LS~2883 if the star was non-rotating. For this reason we fitted a Kurucz stellar atmosphere with a temperature $T_* = 33\,000$~K and an effective gravity $\log g = 4.0$ to the optical data points.  The Kurucz model was fitted to the optical data at frequencies above $10^{13}$~Hz to minimize the influence of the IR excess on the fit. 

Recent observations of Be stars suggest that the circumstellar discs are very thin \citep*[e.g.][]{quirrenbach97,wood97,carciofi11}.  In the SPH modelling presented by \citet{okazaki11} \& \citet{takata12} the disc is assumed to have a temperature of $T_\rmn{disc} = 0.6 \,T_*$ and the height of the disc is assumed to change with radius as $H(R_*)/R_* = 0.024$ (implying $\theta_\rmn{disc}\approx 0.7\degr$).  Therefore, we adopted a disc temperature of $T_\rmn{disc} = 0.6T_*$ and a disc half-opening angle of 1\degr. 

We calculated the Gaunt factors in equation~(\ref{eqn:x_lambda}) assuming a plasma composed of 70 per cent hydrogen (H), 27 per cent helium (He) and 3 per cent carbon, nitrogen and oxygen (CNO).  We assumed that the H and CNO is in an ionized state while the He is neutral.  This was done in order to be consistent with the data tabulated in \citet{waters84}.  

The COG model was fitted to the observational data using the Levenberg-Marquardt method \citep[e.g.][]{press07}.  Only the parameters $n$ and $X_*$ were fitted while $T_\rmn{disc}$, $\theta_\rmn{disc}$ and $R_\rmn{disc}$ were kept at assumed values. The parameters were assumed to have the values  $T_\rmn{disc} = 0.6\,T_* = 19\,800$~K, $\theta_\rmn{disc} = 1$\degr\ and $R_\rmn{disc} = 50R_*$.  

\begin{figure}
 \includegraphics[scale=.65]{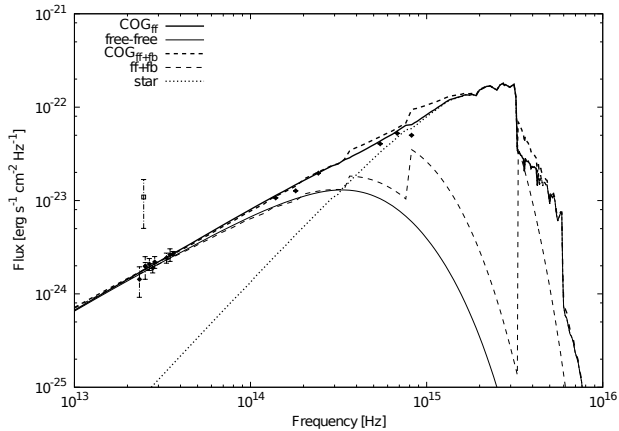}
 \caption{Kurucz atmosphere ($T_*=33\,000$~K, $\log g = 4.0$) and COG method fitted to observational data of LS~2883 (Table~\ref{tab:ls2883_all}). See text for details. } 
 \label{fig:cog_ls2883_fixed}
\end{figure}

Two COG fits were considered, one for free-free+free-bound scattering and the other for only free-free scattering.  Fig.~\ref{fig:cog_ls2883_fixed} shows the stellar atmosphere (dotted line), the predicted IR flux from the disc for free-free (thin solid line) and free-free+free-bound (thin dashed line) scattering, and the total expected flux fitted to the data points in Table~\ref{tab:ls2883_all} (thick solid and dashed lines).  The flux from the MSX Band C (open square), which lies far above the predicted flux, was not included in the calculation of the COG fit as it appears to be significantly inconsistent with other measurements at this frequencies. The reported quality for Band C is 1, indicating a low significance. The Band C measurement therefore represents an upper limit and not a true measurement.  The detection in Band A (8.2~$\mu$m) is consistent with the observations performed with the VLT, and is included in the fit.  The best fit parameters found for the COG method are given in Table~\ref{tab:cog_fig_all}.

\begin{table}
 \centering
 \caption{Parameters of the COG method shown in Fig.~\ref{fig:cog_ls2883_fixed}, for free-free and free-free+free-bound (ff+fb) Gaunt factors, using the observational data in Table~\ref{tab:ls2883_all}.}
  \label{tab:cog_fig_all}
\begin{tabular}{lcc} \hline
Parameter & free-free & free-free+free-bound\\\hline
$T_*$ & 33\,000~K & 33\,000~K \\
$\log g$ & 4.0 & 4.0 \\
$n$ & 3.0550 & 2.8389\\
$\log_{10} X_*$ & 10.245 & 9.9073\\
$R_\rmn{disc}$ & 50~$R_*$ &  50~$R_*$\\
$T_\rmn{disc}$ & 19\,800~K & 19\,800~K\\
$\theta_\rmn{disc}$ & 1\degr & 1\degr\\\hline
\end{tabular}
\end{table}

The calculated free-free+free-bound Gaunt factors create large discontinuities in the predicted flux at frequencies $\ga 2\times10^{14}$~Hz.  At frequencies below this, the IR spectrum is dominated by free-free scattering.  These large discontinuities are not seen in COG fits to other Be stars discussed in \citet{waters86}, as these authors calculated the COG model to a maximum frequency of $\approx 3\times10^{14}$~Hz, which is the limit of the tabulated Gaunt factors in \citet{waters84}.  Below this frequency the influence of the free-bound scattering is still minimal. The free-free scattering produces a smoother curve and this spectrum was adopted for anisotropic modelling IC scattering  from \psrbl.

\subsection{Modelling the IC gamma-ray spectrum}
\label{sec:modelling_ic}

The anisotropic IC scattering rate is determined by performing the quadruple integration in equation~(\ref{eqn:AICtotal}). This integration must be calculated numerically for both the star and the circumstellar disc.  For the star, the photon distribution $n_\rmn{ph}$ is assumed to be a blackbody distribution while for the circumstellar disc the photon distribution is calculated at each point in the orbit using the modified COG method and the best fit free-free parameter to the optical and IR data (Table~\ref{tab:cog_fig_all} and Fig.~\ref{fig:cog_ls2883_fixed}).  

The shape of the electron distribution, $n_\rmn{e}$, depends on the location of the electrons. In the pre--shock region it is assumed that the electron distribution is mono-energetic and only cools through inverse Compton scattering \citep[e.g.][]{kirk99}.  The Lorentz factor of the pre-shock pulsar  wind is typically of the order of $\gamma_p \sim 10^6$, based on modelling of the Crab Nebula, while the possibility of $\gamma_p \sim 10^4$ electrons has been suggested by \citet{khangulyan12} to explain the post-periastron flare observed by {\em Fermi}/LAT after the December 2010 periastron passage.  We have considered these two mono--energetic distributions in the AIC scattering calculation.

In the post--shock region the electrons are re-accelerated to a power law distribution 
\[
  n_\rmn{e}(\gamma)  =  \left\{ \begin{array}{cl}                                                 
K_\rmn{e} \gamma^{-p} & \gamma_\rmn{min} < \gamma < \gamma_\rmn{max}, \\
0 & \rmn{elsewhincluded ere.} \end{array} \right.
\]
where $K_\rmn{e}$ is a constant and $p$ is the electron index.  These post--shock electrons can cool via radiative and adiabatic processes.  
As was shown by \citet{vansoelen11}, if the adiabatically cooled electron spectrum considered by \citet{kirk99} is used, the additional target photons associated with the IR excess result in an increase in the scattering rate at GeV energies.  This electron distribution has an electron index of $p=2.4$ and a distribution lying between $\gamma_\rmn{min} = 5.4\times10^5$ and $\gamma_\rmn{max}=5.4\times10^7$.  We have adopted this distribution to model the IC scattering of post--shock electrons. Additional radiative cooling (e.g.\ synchrotron emission) has not been considered.

It is not an unreasonable assumption that adiabatic cooling dominates near the stand-off shock in \psrb, as the synchrotron radiative cooling time is
\[
t_\rmn{sync} \sim 770 \left(\frac{\gamma}{10^6}\right)^{-1} \left(\frac{B}{1~\rmn{G}}\right)^{-2}~\rmn{sec}, 
\]
which, if the particles are travelling near the speed of light, implies the electrons traverse a distance
$D_\rmn{sync} \sim 36 R_*$ before cooling. This is comparable to the assumed size of the circumstellar disc ($R_\rmn{disc} \approx 50R_*$), therefore near the stand-off shock only adiabatic cooling is dominant. 

\begin{figure}
 \includegraphics[scale=.45]{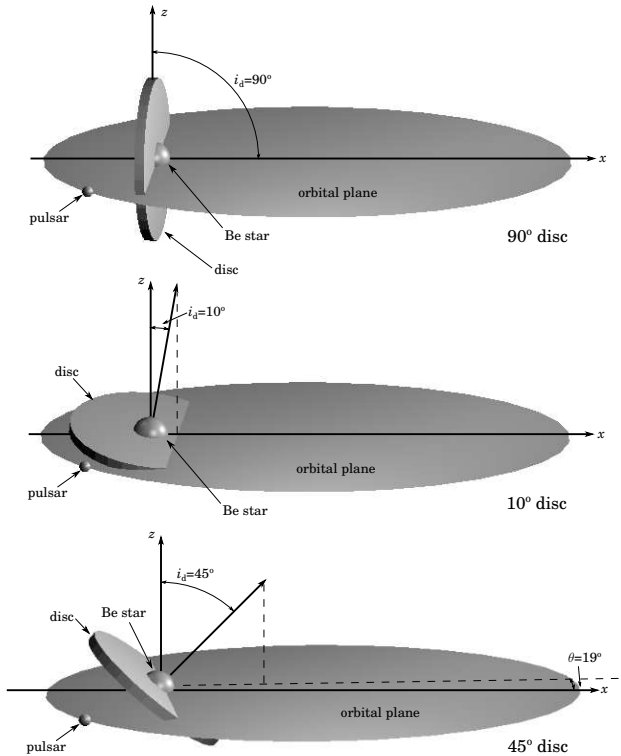}
 \caption{Diagram showing the three disc orientations that we considered in this study. The size of the pulsar, Be star/disc and orbit are not shown to scale.}
 \label{fig:discs}
\end{figure}

We considered three orientations of the circumstellar disc in this study (Fig.~\ref{fig:discs}): the first with the disc plane orientated at 90\degr\ to the orbital plane with the disc normal lying in the orbital plane parallel to the semi-major axis; a second with the disc plane orientated at 10\degr\ to the orbital plane, with the same rotation; and the last using the orientation proposed by \cite{chernyakova06} and modelled by \cite{okazaki11}, with the disc plane orientated at 45\degr\ to the orbital plane and the disc normal rotated 19\degr\ to the semi-minor axis.  Hereafter the orientations will be identified as the 90\degr\ disc, 10\degr\ disc and 45\degr\ disc respectively. For all cases the circumstellar disc is assumed to have a radius $R_\rmn{disc} = 50R_*$. 

For the circumstellar disc, a half-opening angle of $\theta_\rmn{disc}=1$\degr\ was used.  For such a small opening angle the the angle to the normal, $\vartheta$ in equation~(\ref{eqn:AICtotal}), can be approximated by the polar angle, i.e.\ $\vartheta = \theta$.  The disc temperature was taken to be $T_\rmn{disc} = 0.6 T_*$ and the disc was considered to extend to a distance of $R_\rmn{disc} = 50R_*$.  The stellar, orbital and disc parameters are summarized in Table~\ref{tab:disc_parameters_new}.

\begin{table}
 \caption{Stellar and disc parameters used to model LS~2883 in this study.}
\label{tab:disc_parameters_new}
\begin{center}
\begin{tabular}{lccc} \hline
Stellar parameters & × & × & ×\\ \hline
$M_\rmn{star}$ & × & 31~M\sun & ×\\
$R_*$ & × & 9.3~R\sun & ×\\
$M_\rmn{p}$ & × & 1.4~ & ×\\
$T_*$ & × & 33\,000~K & × \\ \hline
Orbital parameters \\ \hline
Inclination & & 33\degr \\ 
Eccentricity & & 0.87 \\
Longitude of periastron & & 138.65\degr \\ \hline 
Disc parameters\\ \hline 
$T_\rmn{disc}$ (K) & & $19\,800$ \\
n & &3.055 \\
$R_\rmn{disc}$ & × & $50~R_\rmn{star}$ & ×\\
$\theta_\rmn{disc}$ & & 1\degr  \\
Tilt to orbital plane & 90\degr & 45\degr & 10\degr\\
Tilt to semi-major axis & 0\degr & 19\degr & 0\degr\\ \hline
\end{tabular}
\end{center}
\end{table}

We considered three electron distributions: a post-shock power law distribution with $p=2.4$ between $5.4\times10^5 < \gamma < 5.4\times10^7$; a pre-shock mono-energetic distribution centred at $\gamma=10^4$; and a pre-shock mono-energetic distribution centred at $\gamma=10^6$.  The IC scattering results from these distributions are discussed below.

\subsubsection{Scattering spectrum  -- post-shock electrons}

\begin{figure}
 \includegraphics[scale=.65]{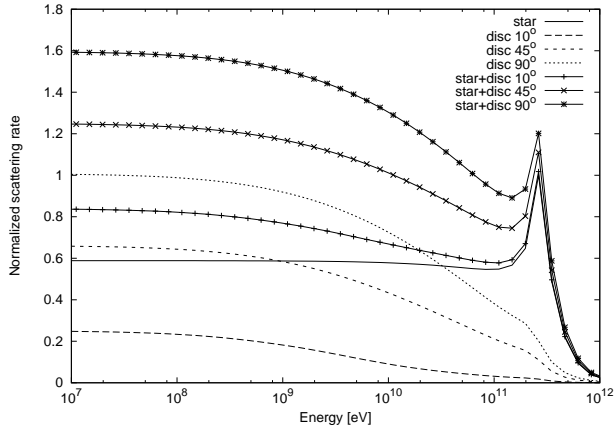}
 \caption{Modelled AIC scattering rate for the post-shock, adiabatically cooled electron distribution at periastron. Shown is the stellar photon contribution (solid line) and the  disc contribution (dashed lines) for the different orientations of the circumstellar disc.  Also shown is the summed contribution from both the star and disc (mark lines). The scattering rate has been normalized to the maximum stellar scattering rate in the figure.}
 \label{fig:scattering_bw}
\end{figure}

The predicted AIC scattering rate, ($dN_\rmn{total}/dt\,d\epsilon_1$), from the post-shocked, adiabatically cooled electron spectrum at periastron is shown in Fig.~\ref{fig:scattering_bw}.  
The solid black line shown in the figure represents the scattering rate due to photons originating from the (unobscured) star.  The dashed lines show the contribution due to the circumstellar disc photons, for the different orientations considered.  Also shown are the combined contributions (star+disc) for the different orientations.
 The scattering rate has been scaled (normalized) to the peak (unobscured) stellar contribution, to illustrate better the ratio between the star and disc contributions. The disc contribution is highest at lower energies, since the scattering occurs in the Thomson limit, while the stellar contribution peaks at a $\rmn{few}\times10^{11}$~eV, due to the scattering occurring mainly in the Klein-Nishina limit.

As is shown in Fig.~\ref{fig:scattering_bw}, the 45\degr\ and 90\degr\ disc contributions exceed the stellar contribution below $\sim 1$ and $\sim 50$~GeV respectively. This effectively means that the IC spectrum at GeV energies is dominated by scattering from the circumstellar disc for these two orientations.  The modelled spectrum shows that the total scattering rate (star+disc contribution) is significantly greater than the stellar contribution alone.  

\subsubsection{Light curve -- post-shock electrons}

\begin{figure}
 \includegraphics[scale=.7]{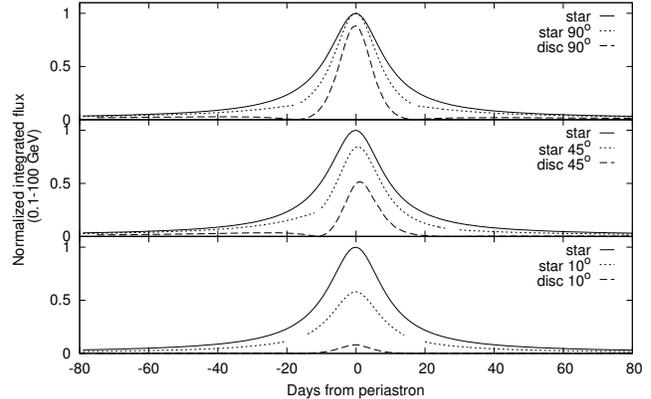}
 \caption{Modelled integrated flux for the AIC scattering of the post-shock, adiabatically cooled electron distribution at {\em Fermi}/LAT energies ($\sim 0.1 - 100$~GeV), normalized to the maximum stellar contribution. For each disc orientation, i.e.\ 90\degr\ (top), 45\degr\ (middle) and 10\degr\ (bottom), the disc contribution is compared with the unobscured and the obscured stellar contribution.}
 \label{fig:aic_light_curve_fermi_new}
\end{figure}

\begin{figure}

 \includegraphics[scale=.7]{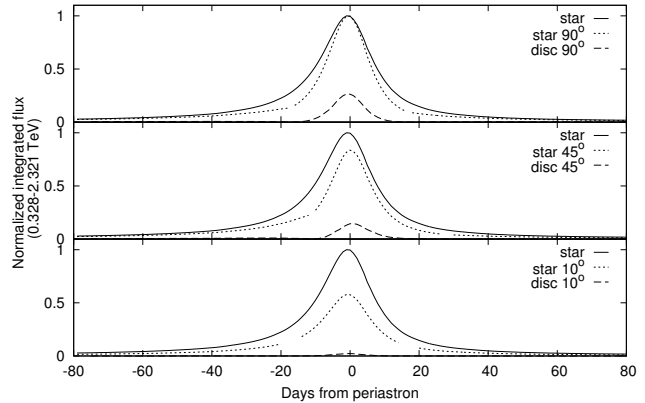}
 \caption{Same as Fig.~\ref{fig:aic_light_curve_fermi_new}, but for the H.E.S.S.\ energy range.}
 \label{fig:aic_light_curve_hess_new}
\end{figure}

The predicted integrated fluxes around periastron, using the parameters above,  are shown in Figs.~\ref{fig:aic_light_curve_fermi_new} \& \ref{fig:aic_light_curve_hess_new} for the {\em Fermi}/LAT and H.E.S.S.\ energy ranges respectively.   The modelled light curves show the integrated flux around periastron, which has been scaled to the maximum (unobscured) stellar contribution.  For the {\em Fermi}/LAT light curve the integration energy range is $0.1 - 100$~GeV, while for the H.E.S.S.\ light curves the energy range is $0.380 - 2.321$~TeV.   The H.E.S.S.\ range was chosen to correspond to the results presented in \citet{aharonian05}.  The lower limit corresponds to their fig.~5, which presents the integrated flux above 380~GeV, while the upper limit, 2.321~TeV, is the highest reported energy in their fig.~4.\footnote{The exact value of the upper limit is given in data released with the publication, available from the H.E.S.S.\ website: \textit{www.mpi-hd.mpg.de/hfm/HESS/pages/publications/auxiliary/AA442\_1.html}.}

The stellar contribution in Fig.~\ref{fig:scattering_bw} is calculated assuming the full solid angle of the star is observable by the pulsar/electron cloud (an unobstructed star,  applicable to no circumstellar disc).  However, if the disc is thick enough to obscure part of the star, the photon contribution from the star will decrease, reducing the stellar contribution to the IC scattering.  Also shown in Figs.~\ref{fig:aic_light_curve_fermi_new} \& \ref{fig:aic_light_curve_hess_new} are the modelled gamma-ray light curves if the disc obscures the star, which changes the shape and decreases the flux contribution from the star around periastron.   The gaps in the modified stellar light curves correspond to periods when the pulsar is passing through the disc.  The light curves show that the disc contribution is a minimum around the disc crossing epoch ($\tau \pm 20$~days) and a maximum around periastron. 

\subsubsection{Scattering spectrum -- pre-shock electrons}

\begin{figure}
 \includegraphics[scale=.65]{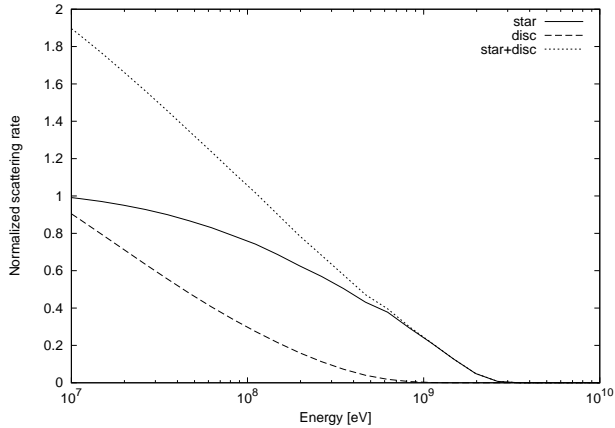}
 \caption{The normalized scattering spectrum for a mono-energetic electron distribution centred at $\gamma = 10^4$. The figure is normalized to the relevant maximum stellar contribution.}
 \label{fig:scattering_four}
\end{figure}

\begin{figure}
 \includegraphics[scale=.65]{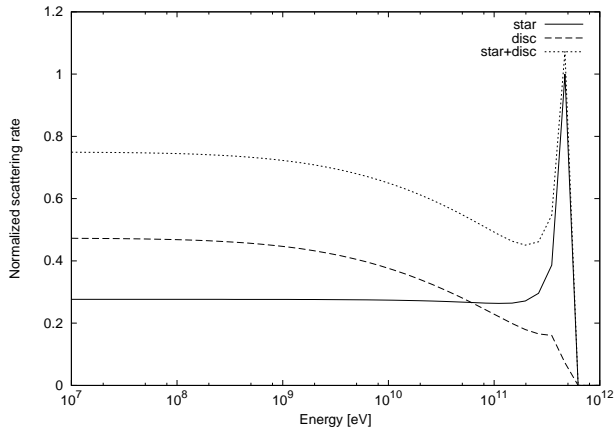}
 \caption{The normalized scattering spectrum for a mono-energetic electron distribution centred at $\gamma = 10^6$. The figure is normalized to the relevant maximum stellar contribution.}
 \label{fig:scattering_six}
\end{figure}

The modelled normalized AIC scattering spectrum for pre-shock electrons with energies $\gamma=10^4$ and $\gamma=10^6$  at periastron are shown in Figs.~\ref{fig:scattering_four} and \ref{fig:scattering_six} respectively.  Only a 90\degr\ disc has been considered for the pre-shock electrons.   The two electron energies which are considered, were suggested by \citet{khangulyan12} ($\gamma=10^4$) and \citet{kirk99} ($\gamma=10^6$).  The $\gamma=10^4$ spectrum (Fig.~\ref{fig:scattering_four}) shows no indication of a peak in the scattering rate, usually associated with scattering in the Klein-Nishina regime.  The highest scattering rate occurs at low energies for both the stellar and disc contributions,  indicating that the scattering occurs in the Thomson limit.  The predicted scattering spectrum for the $\gamma=10^6$ electrons (Fig.~\ref{fig:scattering_six}) shows that the stellar photons scatter in the Klein-Nishina regime (implied by the peak in the scattering rate at $\sim 5\times10^{11}$~eV), while the disc photons scatter in the Thomson regime.  The scattering from the disc is higher than the stellar contribution for energies below $\sim 10^{10}$~eV and the inclusion of the disc contribution increases the total scattering rate by a factor of $\sim2$.

\subsubsection{Lightcurve -- pre-shock electrons}

\begin{figure}
 \includegraphics[scale=.7]{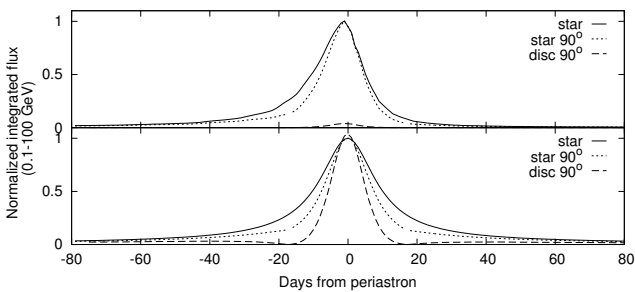}
 \caption{The predicted integrated flux at {\em Fermi}/LAT energies from  the AIC scattering of pre-shocked electrons for $\gamma = 10^4$ (top) and $\gamma=10^6$ (bottom). The figures are normalized to the relevant maximum stellar contribution.}
 \label{fig:fermi_multiplot_preshock}
\end{figure}

The predicted integrated flux at {\em Fermi}/LAT energies around periastron are shown for  $\gamma=10^4$ and $\gamma=10^6$ pre-shock electrons in Fig.~\ref{fig:fermi_multiplot_preshock}, scaled to the maximum stellar contribution.  Only the 90\degr\ disc has been considered, using the same orientation as for the post-shock electrons.  While the $\gamma=10^4$ electrons do not significantly alter the light curve, the $\gamma=10^6$ electrons produce a disc contribution which is significant across the period under consideration, resulting in a higher flux than the stellar contribution at periastron.   The disc minimum occurs around the disc crossing epoch ($\sim \tau\pm 20$~days) while the maximum occurs around periastron. 

\section{Discussion}
\label{sec:discussion}

\subsection{Flux increase}

\begin{figure}
 \includegraphics[scale=.68]{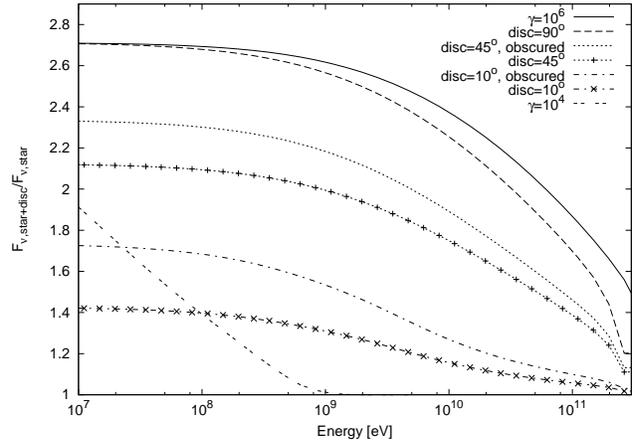}
 \caption{Ratio of the total flux (star+disc contribution) to the flux contribution due to the stellar photons, calculated at periastron.  Indicated on the graph are the adiabatic cooled spectra, calculated for the three disc orientations, with and without the star being obscured.  Also shown are the results for the scattering of pre-shock mono-energetic electrons ($\gamma=10^4$ and $\gamma=10^6$).}
 \label{fig:ratio_peri_neg}
\end{figure}

The predicted increase in the flux at periastron due to the contribution of photons from the disc is shown in Fig.~\ref{fig:ratio_peri_neg} for different electron distributions (pre- and post-shock).  The predicted increase in the flux is of the same order as the isotropic approximation \citep{vansoelen11} and is found to be greater than a factor $\sim 2$ below GeV energies.

Since the anisotropic calculation is highly directional, photons that arrive over the wide solid angle of the disc do not all scatter with the same energy.  Unlike the isotropic approximation where an average scattering angle was considered and the photons were considered to be isotropically distributed, the anisotropic model is based on a photon distribution which is spread out over the solid angle of the disc.  This results in a slight decrease in the influence of the circumstellar disc in the anisotropic calculation.

The increase in the AIC spectrum is of the same order as the isotropic approximation because the IR excess predicted in this paper is higher than considered in \citet{vansoelen11}.  This is due to the change in the colour excess.  A colour excess of E(B-V)$=1.05$ was previously proposed by \citet{johnston94}, however \citet{negueruela11} have proposed a newer value of E(B-V)$=0.85$.  Since the reddening effects are lower, less correction is applied to the observed flux.  Since interstellar reddening influences the optical region far more than the IR region of the spectrum, this results in a lower optical flux relative to the IR flux.  This implies that the IR excess is higher than previously considered.  The disc is therefore brighter because of a higher disc temperature and is brighter relative to the star.  This relative increase in the photon spectrum increases the influence of the disc photon contribution.

\subsection{Light curves}

The predicted AIC, gamma-ray light curves (Figs.~\ref{fig:aic_light_curve_fermi_new}, \ref{fig:aic_light_curve_hess_new} \& \ref{fig:fermi_multiplot_preshock}) show that the disc contribution exhibits three properties: firstly, the contribution is a maximum around periastron; secondly, there is a slight increase just before the disc crossing epoch and a minimum at the disc crossing; and lastly, the rise after the disc crossing is much faster than the slow rise just before the disc crossing.  
This modelled disc contribution light curve is a combination of the photon distribution of the circumstellar disc and the IC scattering angle, as discussed below.

\begin{figure}
 \centering
 \includegraphics[scale=.68]{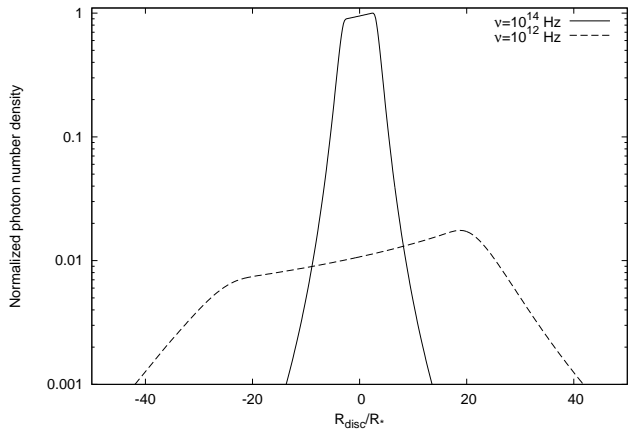}
 \caption{Profile of the normalized photon number density,  at $10^{12}$~and $10^{14}$~Hz,  from the circumstellar disc at $20$ days before periastron plotted through the centre of the disc. Note the y-axis is scaled logarithmically.}
 \label{fig:disc_ir_profile}
\end{figure}

As shown by the profile of the normalized photon number density originating from the disc (Fig.~\ref{fig:disc_ir_profile}), higher frequency photons occur mainly near the centre of the circumstellar disc while longer wavelengths remain constant further out in the disc.
However, the photon number density is much higher at optical frequencies, peaking near $\nu \sim 10^{14}$~Hz (Fig.~\ref{fig:cog_ls2883_fixed}).  These higher frequency photons mainly originate near the centre of the disc, close to the surface of the star.  At the position where the pulsar passes through the disc, $\sim 50 R_*$ from the centre of the disc, the photon density is much lower.  While the target photons from the disc originate from a wide solid angle near the disc crossing epoch, the bright central region only extends over a small solid angle at large polar angles.
Near periastron the pulsar is closer to the star and the bright region of the disc, providing a larger solid angle of brighter photons for IC scattering. This results in a much larger disc contribution towards the AIC scattering at periastron and not at the disc crossing epoch.

The scattering from the disc increases rapidly after the disc crossing epoch due to more favourable scattering angles.  Due to the system orientation, before the disc crossing, all scatterings occur predominantly head-to-tail ($\theta_\rmn{sc} =0\degr$) while after the disc crossing predominantly head-to-head ($\theta_\rmn{sc} = 180\degr$) collisions occur.  On the other hand, the AIC scattering of the stellar photons follows a smooth transition which results in a smooth light curve and not the sharp increase and decrease seen for the disc contribution.  

The combined effects of the photon density and the IC scattering angle result in the predicted AIC light curves.  Additional effects, such as the heating of the circumstellar disc due to the pulsar crossing or changes in the disc size near periastron, have not been considered.

It may be possible that shock heating of the circumstellar disc could locally increase the disc intensity, providing additional photons for IC scattering, and significantly influence the gamma-ray light curve. \citet{khangulyan12} have suggested that the observed {\em Fermi}/LAT flare is due to the shock heating of the disc and the IC scattering of pre-shocked pulsar wind with a Lorentz factor of $\gamma\sim10^4$.  However, the model proposes a large opening angle for the circumstellar disc and requires the pulsar to leave the disc at $\sim\tau+30$~days near the observed {\em Fermi}/LAT flare.  This is in contradiction to the radio observations which detect the pulsed radio emission $\tau+15$~days after periastron \citep{abdo11}.

\subsection{Consequences of the model}

By considering only the IR excess from the circumstellar disc and not considering any of the other consequences of changes in the density, the size of the pulsar wind nebula with orbital motion or radiative cooling effects, the total gamma-ray flux increases at GeV energies by a factor $\sim 2$. While it was thought that the extended size of the disc may completely dominate at the disc cross-epoch, it is shown by the modified COG method that the IR intensity at the distance of the disc crossing is  faint and the majority of the emission occurs close to the central star.  

\begin{figure}
 \centering
 \includegraphics[scale=.65]{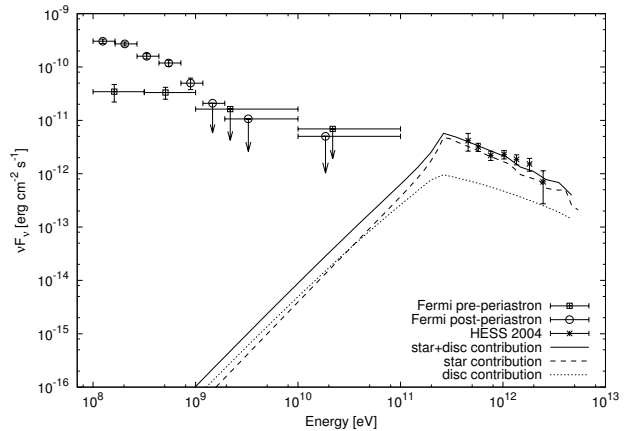}
 \caption{Comparison of the predicted model to the {\em Fermi}/LAT and H.E.S.S.\ observations. The vertical arrows indicate upper limits on the {\em Fermi}/LAT detections.}
 \label{fig:fermi_hess_model}
\end{figure}

Fig.~\ref{fig:fermi_hess_model} shows the calculated gamma-ray energy spectrum at periastron in comparison to the H.E.S.S.\ and {\em Fermi}/LAT observations.  The model curve has been been linearly scaled to the observations.
The predicted gamma-ray flux lies orders of magnitudes below the {\em Fermi}/LAT detection limit and is therefore undetectable with current gamma-ray telescopes. At Very High Energies (VHE), while the flux is high enough for detection, the predicted variability which would result from a change in the disc IR contribution is smaller than the error in the H.E.S.S.\ measurements and therefore would not be detected at a significant level. However, Be stars are also known to be variable at optical frequencies.  If the optical magnitude of the system changes significantly, this could result in a higher variability at TeV energies.
 
\subsection{State of the disc post-periastron}
\label{sec:state_post_periastron}

SPH simulations \citep{okazaki11,takata12} of \psrbl\ around periastron have suggested that the disc could be disrupted by the passage of the pulsar during the periastron passage.  As shown by these authors, the degree to which the disc is disrupted depends on the density of the disc.  For a larger density ($\rho_0 \sim 10^{-9}$~g~cm$^{-3}$) the disc remains more stable.   This could result in changes in the observed optical  and IR emission from the disc.   However, observations near periastron on 5 January 2011 with VISIR/VLT  found a flux that was of the same order as earlier Band A measurements with the \emph{Midcourse Space Experiment} (MSX) (Table~\ref{tab:ls2883_all}). 
Similarly, spectroscopic observations of LS~2883 with the SAAO 1.9m telescope, taken four months after periastron show a very prominent H${\alpha}$ emission line and no indication of stellar  absorption, indicating that the circumstellar disc completely dominates the stellar absorption at the same frequency.  These observations, of the IR excess and H$\alpha$ emission line, which took place after periastron, show that the Be star is still surrounded by a substantial circumstellar disc.  The observed equivalent width for the H$\alpha$ line is EW~$\approx -50\pm5$~\AA, very similar to the measurement reported by \citet{negueruela11} (EW~$=-54\pm2$~\AA). This may be an indication that the disc remains prominent despite the pulsar passage or, alternatively, may indicate a very rapid recovery.  Similarly the observations obtained with SALT during May 2012, far away from periastron, still show a prominent H$\alpha$ line.

\subsection{Connection with the {\em Fermi}/LAT flare}

The flare detected with {\em Fermi}/LAT after the last periastron passage was unexpected.  We have suggested that the IR excess associated with the circumstellar disc can have a large influence on the gamma-ray production, increasing the flux by a factor $\sim2$.  However this increase is too low to explain the {\em Fermi}/LAT flare.  As is shown by the COG model, the majority of the IR emission is produced close to the stellar surface and not near the location of the pulsar-disc crossing.  This results in the majority of the disc contribution occurring close to periastron and not near the disc crossing phase.  It has been suggested by \citet{khangulyan12} that the pre-shock scattering of $\gamma=10^4$ electrons could cause the observed {\em Fermi}/LAT flare if the photon density from this disc is large enough.  These authors have suggested that a luminosity of 40 per cent of the stellar luminosity is required, and that the additional photons must be originate over a large region. However, the observations undertaken during and after the previous periastron passage do not seem to suggest such a large heating effect.  The IR emission from the circumstellar disc remains fairly consistent with the \textit{2MASS} observations, while the H$\alpha$ observations four months after periastron is consistent with earlier observations reported by \citet{negueruela11}.    These observations then suggest that the shock heating which should occur during the pulsar-disc crossing will be more localised around the pulsar.  

\citet*{zabalza11} discussed thermal X-ray emission in gamma-ray binaries as a result of the shock heating of the stellar wind. These authors estimated the total luminosity of the thermal X-rays as some fraction of the total kinetic luminosity of the stellar wind.  We can make a similar approximation of the total amount of energy available for optical/IR emission from shock heating by estimating the kinetic luminosity of the stellar wind from LS~2883. The total kinetic luminosity can be approximated by $L_\rmn{kin} \sim \dot{M} v^2$, where $\dot{M}$ is the mass loss rate, and $v$ is the speed of the wind. Similarly to \cite{kong11} we can estimate the kinetic luminosity for the polar wind to be,
\[
 L_\rmn{kin}^\rmn{pole} \sim 6.3\times10^{33} \left(\frac{\dot{M}}{10^{-8}\rmn{M}_\odot\rmn{yr}^{-1}}\right)\left(\frac{v}{10^8\rmn{cm~s}^{-1}}\right)^2\rmn{erg~s}^{-1}
\]
and for the equatorial wind to be
\[
 L_\rmn{kin}^\rmn{disc} \sim 6.3\times10^{30} \left(\frac{\dot{M}}{10^{-7}\rmn{M}_\odot\rmn{yr}^{-1}}\right)\left(\frac{v}{10^6\rmn{cm~s}^{-1}}\right)^2\rmn{erg~s}^{-1}.
\]
The kinetic luminosity of the circumstellar disc can also be estimated by considering the disc to be Keplerian, with a velocity $v = (G M_*/r)^{1/2}$.  If the density through the circumstellar disc is given by equation~\ref{eqn:rho_distance}, the mass flow rate of the disc through an area d$A$ is given by $\rmn{d}M/\rmn{d}t \sim \rho\, v \rmn{d}A$, where the density and velocity depend on the distance $r$ from the centre of the star.  Integrating over the area of the disc perpendicular to the direction of the disc motion (d$A=2r \tan \theta_\rmn{disc} \rmn{d}A$) gives the total kinetic luminosity as
\begin{eqnarray*}
 L_\rmn{kin}^\rmn{disc} &\sim& \int_{R_*}^{R_\rmn{disc}} \rho(r) \left(\frac{G M_*}{r}\right)^{3/2} 2r \tan \theta_\rmn{disc} \, \rmn{d}r \\ &\sim& 9\times10^{35}~\rmn{erg~s}^{-1}.
\end{eqnarray*}
Here $\rho$ has been calculated from equation~\ref{eqn:rho_distance} and the fitted COG parameters have been used.  Since only a fraction of the available kinetic luminosity can be directed into shock heating, it seems likely that any increase in the IR/optical emission will have a luminosity $\ll 10^{36}$~erg~s$^{-1}$.

It has also been shown by \citet{kong12} that Doppler boosting of the synchrotron emission could produce the required energy level to explain the {\em Fermi}/LAT flare.  The resulting flare occurs because of the preferential direction of the shock outflow after the second disc crossing which results in bulk Doppler boosting of material in the cone of the pulsar wind shock front.  This suggests that the {\em Fermi}/LAT flare is the result of the combined effects of shock heating and Doppler shifting.  The SPH simulations undertaken by \citet{takata12} show that circumstellar disc can be deformed around the pulsar, creating a cavity-like structure around the pulsar.  Such a shock heated cavity may then allow for additional photons, which will allow for head-to-head gamma-ray production after the second disc crossing.  

Dedicated optical and IR campaigns around future disc crossing phases are required to investigate the degree to which the circumstellar disc is influenced by the pulsar and pulsar wind.

\section{Conclusion}
\label{sec:conclusion}

The binary star system \psrbl\ has remained a fascinating source since the detection of the pulsar approximately two decades ago.  The interaction between the Be star and the pulsar produces multi-wavelength emission from radio up to TeV gamma-rays, covering 19 decades in energy.   The optical companion in the system is a Be type star, known to possess a circumstellar disc which produces an IR flux.  This flux is observed as an IR excess which is higher than predicted by blackbody and stellar atmosphere models.  Since the gamma-rays in the system are produced by the up-scattering of photons, the presence of an extended and additional photon source could potentially have a significant effect on the gamma-ray production in \psrbl.  

We modified the COG method, first proposed by \citet{waters84}, which allowed us to predict the flux originating from the circumstellar disc at any position near the disc.  This method has been used to predict the photon density as observed from the point-of-view of the pulsar.  The COG method was constrained using optical and IR data obtained from literature and VLT observations.  The photon density was calculated around periastron and we applied it to the problem of AIC scattering. 

The anisotropic scattering considered the change in the photon distribution during the orbit, the variation in the solid angle subtending the disc, and the direction of scattering.  The variation in the gamma-ray spectrum around periastron was calculated for different disc orientations.  

In this study the leptons in the pulsar wind are assumed to be mono-energetic before the stand-off shock and a power-law energy distribution in the post-shock region.  Both a pre-- and post--shock distribution have been considered. For \psrb, it can be shown that adiabatic cooling dominates near the stand-off shock and, for this reason, only adiabatic cooling was considered for the AIC modelling.  

The IC modelling showed that the inclusion of the IR excess associated with the circumstellar disc increased the GeV gamma-rays by a factor $\ga 2$.  This relative increase in the flux is partly dependent on the choice of the electron distribution. For a wide distribution (e.g.\ $10^4 \leq \gamma \leq 10^7$), as was shown for the isotropic modelling \citep[fig.~3 in][]{vansoelen11}, or an electron distribution centred around $\gamma \sim 10^4$ (Fig.~\ref{fig:fermi_multiplot_preshock}), the IC scattering is dominated by the scattering of stellar photons in the Thomson limit.  However, this study has shown that for the electron distribution proposed by \cite{kirk99}, which has been shown to provide a reasonable fit to the 2004 H.E.S.S.\ observations, the IR excess from the circumstellar disc increases the gamma-ray flux by a factor $\geq 2$ at GeV energies.

The predicted light curves show (Figs.~\ref{fig:aic_light_curve_fermi_new}, \ref{fig:aic_light_curve_hess_new} \& \ref{fig:fermi_multiplot_preshock}) that the circumstellar disc is most influential close to periastron and, generally, after the disc crossing. This is a result of the disc being brightest near the stellar surface  and the associated favourable scattering angles after the disc crossing.  While it was thought the predicted gamma-ray flux may be highest near the disc crossing epoch ($\sim \tau \pm 20$~days), the low intensity of the disc at $\sim 50$ stellar radii from the star's surface does not provide a sufficient number of photons to increase significantly the gamma-ray flux in this region.  This study has only considered a steady state disc and has not considered what effects the pulsar passage may have introduced on the circumstellar disc.  Shock heating of the disc or clumping of material may result in regions of higher intensity which could result in more localised flux increases.  

The additional effects such as changes in the electron cooling have not been considered, as these have been addressed by other authors.  This study has focussed on the effect of the IR excess from the circumstellar disc in order to determine its influence on the production of GeV energy gamma-rays.  The study shows that the gamma-ray flux could be increased by a factor greater than two at GeV energies, which is a significant increase.  However, the predicted gamma-ray flux lies far below that detected by {\em Fermi}/LAT  at the previous periastron passage. A combination of a number of effects, including e.g.\ Doppler boosting, shock heating and changes in the shock front are required to explain the flare that was observed by {\em Fermi}/LAT  approximately one month after periastron.

\section*{Acknowledgements}

The authors are very grateful to the late O.C.~de~Jager for suggesting this study. Numerical modelling was undertaken using the University of the Free State's High Performance Computing division and the authors are particularly grateful for the help of A.\ van Eck.  
This research was funded by the South African Square Kilometre Array Project.
Some of the observations reported in this paper were obtained with the Southern African Large Telescope (SALT).
This paper uses observations made at the South African Astronomical Observatory (SAAO).  
This publication makes use of data products from the Two Micron All Sky Survey, which is a joint project of the University of Massachusetts and the Infrared Processing and Analysis Center/California Institute of Technology, funded by the National Aeronautics and Space Administration and the National Science Foundation.
This research made use of data products from the Midcourse Space Experiment. Processing of the data was funded by the Ballistic Missile Defense Organization with additional support from NASA Office of Space Science. This research has also made use of the NASA/ IPAC Infrared Science Archive, which is operated by the Jet Propulsion Laboratory, California Institute of Technology, under contract with the National Aeronautics and Space Administration.
The authors would like to thank the anonymous reviewer for their very helpful comments which have improved this study.

\appendix

\section{Additional constraints on the COG method}
\label{app:limits}

The specific intensity which orientates along any line-of-sight in the disc can be calculated by considering an impact vector $\bmath{q}$ relative to the disc. In Fig.~\ref{fig:cog_fig_paper}, the origin of the coordinate system is at $O$ (the centre of the pulsar), the unit vector $\bmath{\hat{e}_*}$ points towards the centre of the disc (and star) and $\bmath{\hat{e}_{\theta,\phi}}$ in the direction of the incoming photon/line-of-sight.  The value of $q$ is then defined as 
\[
 q = \bar{r} \sin \xi,
\]
where $\bar{r}$ is the radial distance in units of $R_*$ and $\xi$ is defined by 
\begin{equation}
 \cos \xi = \bmath{\hat{e}_*} \cdot \bmath{\hat{e}_{\theta,\phi}}. 
 \label{eqn:xi}
\end{equation}

Any line-of-sight in a direction defined by $\theta$, $\phi$ intercepts the disc mid-plane at a position (Figs.~\ref{fig:cog_fig_paper}  \& \ref{fig:disc-geometry})
\begin{eqnarray*}
x_d & = & H \tan \theta \cos \phi + L,\\
y_d & = & H \tan \theta \sin \phi, 
\end{eqnarray*}
where $x_d$ and $y_d$ are measured relevant to the disc mid-plane such that the origin is in the centre of the star/disc and the $x_d$ and $y_d$ are measured in the plane of the disc.  The height of the disc at any point in this coordinate system is 
\[
 h_d = \sqrt{x^2 + y^2} \tan \theta_\rmn{disc}, 
\]
where $\theta_\rmn{disc}$ is the opening angle of the disc. For any line-of-sight ($\bmath{\hat{e}_{\theta,\phi}}$), we can construct a distance $h_l$ which is perpendicular to the disc mid-plane and intercepts the line-of-sight.  If $h_l$ intercepts the disc mid-plane at any distance $\Delta x$ and $\Delta y$ from the point of the disc plane intercept ($x_d,y_d$), then the length along $h_l$ between the disc mid-plane and the line-of-sight intercepts is 
\begin{equation}
  h_l = \sqrt{\Delta x^2 + \Delta y^2} \tan \theta_k, 
\label{eqn:h_l}
\end{equation}
where $\theta_k = \frac{\pi}{2} - \theta$. To find the point of the disc intercept $\Delta x$ and $\Delta y$ can be solved from $h_l = h_d$, and therefore
\begin{eqnarray}
\nonumber \sqrt{\Delta x^2 + \Delta y^2} \tan \theta_k, & = & \sqrt{x^2 + y^2} \tan \theta_\rmn{disc}\\
 \nonumber &=& \sqrt{(x_d + \Delta x)^2 + (y_d + \Delta y)^2} \tan \theta_\rmn{disc}.
\end{eqnarray}
Since in the plane of the disc, $\Delta y = \Delta x \tan \phi$ the equation above can be solved for $\Delta x$ and hence $\Delta y$ can be calculated.  The solution gives two values for both $\Delta x$ and $\Delta y$, which lie below and above the disc plane.  Special cases occur when $\Delta x$ or $\Delta y$ are equal to zero. The correct solution for the intercept above and below the plane is determined by the direction of the  line-of-sight and its orientation to the disc.  The values of the ($x_1,y_1$) and ($x_2, y_2$) coordinates are then given by
\begin{eqnarray*}
 x_{1,2} &=& x_d + \Delta x_{1,2}\\ 
 y_{1,2} &=& y_d + \Delta y_{1,2}, 
\end{eqnarray*}
and the z coordinates are calculated by 
\begin{eqnarray*}
 z_1 &=& -\left|\sqrt{x_1^2 + y_1^2}\right| \tan \theta_\rmn{disc} \\
 z_2 &=& +\left| \sqrt{x_2^2 + y_2^2}\right| \tan \theta_\rmn{disc}. 
\end{eqnarray*}

Using the solution to the disc intercepts, we then calculate the angles $X_1$ and $X_2$ in Fig.~\ref{fig:cog_fig_paper}.  The two disc intercepts lie at  $(x_1,y_1,z_1)$ and $(x_2,y_2,z_2)$ while the centre of the electron cloud/pulsar lies at $(x=L, y = 0, z = H)$.  The angles $X_1$ and $X_2$ are the angles between the disc intercepts $(x_1,y_1,z_1)$ \& $(x_2,y_2,z_2)$ and the centre of the electron cloud, and are calculated from
\begin{eqnarray*}
  \cos X_1 &=& \frac{(L, 0, H) \cdot (x_1,y_1,z_1)}{ \| (L, 0, H)\| \|(x_1,y_1,z_1)\|},\\
 \cos X_2 &=& \frac{(L, 0, H) \cdot (x_2,y_2,z_2)}{\|(L, 0, H)\| \|(x_2,y_2,z_2)\|}.
\end{eqnarray*}

The integration limits on $w$ in equation~(\ref{eqn:tau}) are then given by
\begin{eqnarray*}
 w_1 &=& \frac{\pi}{2} - \xi - X_1 \\
 w_2 &=& \frac{\pi}{2} - \xi - X_2, 
\end{eqnarray*}
where $\xi$ is defined in equation~(\ref{eqn:xi}).

\end{document}